\documentclass[preprint,aps,prd,nofootinbib]{revtex4}
\usepackage{amsmath}
\usepackage{graphicx}
\usepackage{subfigure}

\newcommand{\bea}{\begin{eqnarray}}
\newcommand{\eea}{\end{eqnarray}}
\newcommand{\beq}{\begin{equation}}
\newcommand{\eeq}{\end{equation}}
\newcommand{\nn}{\nonumber}

\def\/{\over}

\begin{document}

\title{ Fulling-Davies-Unruh effect and spontaneous excitation of an accelerated atom interacting with a quantum scalar field
}
\author{ Zhiying Zhu${^2}$ and Hongwei Yu$^{1,2,}$\footnote{Corresponding author} }
\affiliation{ $^1$CCAST(World Lab.), P. O. Box 8730, Beijing,
100080, P. R. China\\
$^2$Department of Physics and Institute of  Physics,\\
Hunan Normal University, Changsha, Hunan 410081, China
\footnote{Mailing address} }

\begin{abstract}

 We investigate, from the point of view of a coaccelerated
 frame, the spontaneous excitation of  a uniformly accelerated two-level atom
interacting with a scalar field in a thermal state at a finite
temperature $T$  and show that the same spontaneous excitation rate
for the uniformly accelerated atom in the Minkowski vacuum obtained
in the inertial frame can only be recovered in the coaccelerated
frame assuming a thermal bath at the Fulling-Davies-Unruh
temperature $T_{FDU}=a/2\pi$  for what appears to be the Minkowski
vacuum to the inertial observer. Our discussion provides another
example of a physical process different from those examined before
in the literature to better understand the Fulling-Davies-Unruh
effect.

\end{abstract}

\maketitle

\baselineskip=16pt

\section{introduction}

It is well known that the particle content of a quantum field theory
is observer dependent. According to the works of Fulling, Davies and
Unruh (FDU), for a uniformly accelerated observer, the Minkowski
vacuum is seen to be equivalent to a thermal bath of Rindler
particles at a temperature $T_{FDU}=a/2\pi$ (FDU effect \cite{FDU}).
The existence of the FDU effect has been shown to be mandatory for
the consistency of quantum field theory in various cases, where it
has been demonstrated that the way to obtain agreement between
physical observables calculated from the inertial and coaccelerated
frame point of view is by assuming the FDU effect
\cite{Unruch84,Matsas01,Higuchi92,Danilo, Hai Ren}. For the weak
decay of a uniformly accelerated proton, the same lifetime can be
obtained in the inertial frame and the coaccelerated frame
considering the FDU effect \cite{Matsas01}. For the bremsstrahlung
effect associated with a uniformly accelerated point charge,
Higuchi, Matsas and Sudarsky \cite{Higuchi92} have shown that the
same response rate of the uniformly accelerated point charge to the
Larmor radiation can be obtained in both the inertial and
coaccelerated frame assuming the FDU thermal bath with an infinite
number of zero-energy Rindler photons in the Rindler wedge, and the
same has been shown to be true in the presence of boundaries
\cite{Danilo}. Meanwhile the generalization to  the quantum
bremsstrahlung effect for a uniformly accelerated point scalar
source has also been made\cite{Hai Ren}. As another example, besides
the weak decay of the proton and the bremsstrahlung effect,
demonstrating  the necessity of the FDU effect for the consistency
of quantum field theory, we will compute, from the point of view of
a coaccelerated frame, the spontaneous excitation rate of a
uniformly accelerated two-level atom interacting with a massless
scalar field, and show that equality can only be established between
the excitation rates obtained in the inertial and the coaccelerated
frames by assuming the FDU effect.

Let us note that the spontaneous excitation of a uniformly
accelerated atom interacting with a massless scalar field
\cite{Audretsch94,H. Yu05} and with electromagnetic fields
\cite{ZYL06} in a Minkowski vacuum has already been studied in the
inertial frame \cite{Audretsch94} using the formalism proposed by
Dalibard, Dupont-Roc and Cohen-Tannoudji
\cite{Dalibard82,Dalibard84}, which demands a symmetric operator
ordering of atom and field variables and allows one to separately
calculate the contributions of vacuum fluctuations and radiation
reaction to the spontaneous excitation rate of the accelerated atom.
It is found that for a uniformly accelerated atom transition from
ground state to excited states becomes possible even in vacuum. This
phenomenon provides a physically appealing interpretation of the FDU
effect, since it gives a transparent illustration for why an
accelerated detector clicks (See Ref.~\cite{TP85} for a discussion
in a different context and Ref.~\cite{Hu} for  a non-perturbative
approach to study the interaction of a uniformly accelerated
detector, modelled by a harmonic oscillator which may be regarded as
a simple version of an atom, with a quantum field in (3+1)
dimensional spacetime).

 The purpose of the present paper is to show that
the rate of change of the atomic energy for a uniformly
accelerated atom interacting with a massless scalar field in
vacuum obtained in the inertial frame for both cases with and
without boundaries \cite{Audretsch94, H. Yu05} can be recovered in
the coaccelerated frame by using Fulling's quantization in
conjunction with the fact that the Minkowski vacuum is a thermal
state of Rindler particles at a temperature $T_{FDU}=a/2\pi$. So,
we will consider a uniformly accelerated two-level atom
interacting with a massless scalar field in the coaccelerated
frame both in space with and without the presence of boundary, and
evaluate the rate of change of the mean atomic energy of the atom
assuming a thermal bath of Rindler particles at a finite
temperature $T$. Since the rate of change of the atomic energy is
a scalar, the same value of this observable must be obtained in
the inertial and coaccelerated frame. We will show that only when
the temperature equals the  FDU value, i.e., $T=a/2\pi$, can
agreement in both frames be obtained.

\section{The general formalism for thermal fluctuations and radiation
reaction}

We consider a two-level atom in interaction with a massless scalar
field, and assume the atom has a constant proper acceleration $a$
along the $x$ direction in the Minkowski coordinates. In the Rindler
wedge, the line element \cite{QFT} can be described by
\begin{eqnarray}
ds^2=e^{2a\xi}(d\tau^2-d\xi^2)-dy^2-dz^2\;.
\end{eqnarray}
The Rindler coordinates are related to the usual Minkowski
coordinates by
\begin{eqnarray}
t={e^{a\xi}\/a}\sinh{a\tau}\;,\ \ x={e^{a\xi}\/a}\cosh{a\tau}\;.
\end{eqnarray}
In the Rindler coordinates our accelerated atom is stationary on the
trajectory $x=(\tau,\xi,y(\tau),z(\tau))$. The stationary trajectory
guarantees that the undisturbed atom has stationary states,
$|-\rangle$ and $|+\rangle$, with energies $-{1\/2}\omega_0$ and
$+{1\/2}\omega_0$ and a level spacing $\omega_0$. The Hamiltonian
that governs the time evolution of the atom with respect to the
proper time $\tau$ is written in Dick's \cite{Dicke} notation
\begin{eqnarray}
H_A(\tau)=\omega_0R_3(\tau)\;,
\end{eqnarray}
where $R_3={1\/2}|+\rangle\langle+|-{1\/2}|-\rangle\langle-|$. The
free Hamiltonian of the quantum scalar field with respect to $\tau$
is
\begin{eqnarray}
H_F(\tau)=\int d\omega\int dk_y\int
dk_z~\omega~b_{\omega,k_y,k_z}^\dag(\tau)
~b_{\omega,k_y,k_z}(\tau)\;,
\end{eqnarray}
where $b_{\omega,k_y,k_z}^\dag$ and $b_{\omega,k_y,k_z}$ denote the
creation and annihilation operators for a Rindler particle with
transverse momentum $k_y$ and $k_z$ and frequency $\omega$.
 The quantization of the field in the Rindler
wedge can be  carried out by the expansion in terms of annihilation
and creation operators as
\begin{eqnarray}
\phi(x)=\int d\omega\int dk_y\int
dk_z[\,b_{\omega,k_y,k_z}(\tau)\upsilon_{\omega,k_y,k_z}(x)+H.c.]\;,\label{phi}
\end{eqnarray}
 where $\upsilon_{\omega,k_y,k_z}$ are the Rindler modes.  Let us note
that the positive acceleration $a$ of the atom makes only the
Rindler wedge $R^+$ ($x>|t|$) completely accessible to a
coaccelerated observer \cite{Boulware}. Thus in the following
sections, we will only consider the Rindler modes in $R^+$.
Following Ref.~\cite{Audretsch94}, the Hamiltonian that describes
the interaction between the atom and the quantum field can be
written
\begin{eqnarray}
H_I(\tau)=\mu R_2(\tau)\phi(x)\;.
\end{eqnarray}
Here $\mu$ is a coupling constant which we assume to be small, and
$R_2={1\/2}i(R_--R_+)$, where $R_+=|+\rangle\langle-|$ and
$R_-=|-\rangle\langle+|$. The coupling is effective only on the
trajectory  of the atom. The Heisenberg equations of motion for the
dynamical variables of the atom and the field can be derived from
the Hamiltonian $H=H_A+H_F+H_I$:
\begin{eqnarray}
{d\/d\tau}R_\pm(\tau)=\pm
i\omega_0R_\pm(\tau)+i\mu\phi(x)[R_2(\tau),R_\pm(\tau)]\;,
\end{eqnarray}
\begin{eqnarray}
{d\/d\tau}R_3(\tau)=i\mu\phi(x)[R_2(\tau),R_3(\tau)]\;,
\end{eqnarray}
\begin{eqnarray}
{d\/d\tau}b_{\omega,k_y,k_z}(\tau)=-i\omega
b_{\omega,k_y,k_z}(\tau)+i\mu
R_2(\tau)[\phi(x),b_{\omega,k_y,k_z}(\tau)]\;.
\end{eqnarray}
We can split the solutions of these equations of motion into the
``free" and ``source" parts.

Let us now assume that the system is in a thermal bath of arbitrary
temperature $T$ as seen by the coaccelerated observer (Rindler
observer), so that the scalar field is in a thermal state (as
opposed to a vacuum state in the inertia frame) and the density
matrix is given by $\rho=e^{-\beta H_F}=e^{-H_F/T}$, and the atom is
in the state $|a\rangle$. Our aim is to identify and separate the
contributions of quantum thermal fluctuations (as opposed to vacuum
fluctuations in the inertial frame) and radiation reaction to the
rate of change of the mean atomic energy. For this purpose, we
choose a symmetric ordering between the atom and field variables,
and separate the two contributions of thermal fluctuations and
radiation reaction to the rate of change of $H_A$ ( cf. Ref.
\cite{Audretsch94,Dalibard82,Dalibard84}),
\begin{eqnarray}
{dH_A(\tau)\/d\tau}=\biggl({dH_A(\tau)\/d\tau}\biggr)_{TF}+\biggl({dH_A(\tau)\/d\tau}\biggr)_{RR}\;,
\end{eqnarray}
where
\begin{eqnarray}
\biggl({dH_A(\tau)\/d\tau}\biggr)_{TF}={1\/2}i\omega_0\mu(\phi^f(x)[R_2(\tau),R_3(\tau)]+
[R_2(\tau),R_3(\tau)]\phi^f(x))\;,
\end{eqnarray}
representing the contribution of thermal fluctuations and
\begin{eqnarray}
\biggl({dH_A(\tau)\/d\tau}\biggr)_{RR}={1\/2}i\omega_0\mu(\phi^s(x)[R_2(\tau),R_3(\tau)]+
[R_2(\tau),R_3(\tau)]\phi^s(x))\;,
\end{eqnarray}
representing that of radiation reaction.

We can separate $R_2$ and $R_3$ into their free part and source
part, and take the expectation value in the field's and atom's
states. We obtain, up to order $\mu^2$,
\begin{eqnarray}
\biggl\langle{dH_A(\tau)\/d\tau}\biggr\rangle_{TF}^R=2i\mu^2\int_{\tau_0}^\tau
d\tau'C^F_\beta(x,x'){d\/d\tau}\chi^A(\tau,\tau')\;,\label{tf}
\end{eqnarray}
\begin{eqnarray}
\biggl\langle{dH_A(\tau)\/d\tau}\biggr\rangle_{RR}^R=2i\mu^2\int_{\tau_0}^\tau
d\tau'\chi^F_\beta(x,x'){d\/d\tau}C^A(\tau,\tau')\;,\label{rr}
\end{eqnarray}
where the superscript $R$ denotes  the calculation performed in the
Rindler coordinates and $| \rangle = |a,\beta \rangle$ representing
the atom in the state $|a\rangle$ and the field in the  thermal
state $|\beta \rangle$. The statistical functions $C^F_\beta(x,x')$
and $\chi^F_\beta(x,x')$ of the  field at a nonzero temperature
$T=1/\beta$ are defined as
\begin{eqnarray}
C^F_\beta(x,x')={1\/2}\langle\{\phi^f(x),\phi^f(x')\}\rangle_\beta
={1\/2}\textmd{tr}(\rho\{\phi^f(x),\phi^f(x')\})/\textmd{tr}(\rho)\;,\label{ct}
\end{eqnarray}
\begin{eqnarray}
\chi^F_\beta(x,x')={1\/2}\langle~[\phi^f(x),\phi^f(x')]~\rangle_\beta
={1\/2}\textmd{tr}(\rho[\phi^f(x),\phi^f(x')])/\textmd{tr}(\rho)\;,\label{xt}
\end{eqnarray}
and those of the atom as
\begin{eqnarray}
C^A(\tau,\tau')={1\/2}\langle
a|\{R_2^f(\tau),R_2^f(\tau')\}|a\rangle\;,
\end{eqnarray}
\begin{eqnarray}
\chi^A(\tau,\tau')={1\/2}\langle
a|~[R_2^f(\tau),R_2^f(\tau')]~|a\rangle\;.
\end{eqnarray}
$C^F_\beta$ ($C^A$) is called the symmetric correlation function of
the scalar field at a finite temperature $T=1/\beta$ (atom),
$\chi^F_\beta$ ($\chi^A$) its linear susceptibility. The explicit
forms of the statistical functions of the atom are given by
\begin{eqnarray}
C^A(\tau,\tau')={1\/2}\sum_b|\langle a|R_2^f(0)|b\rangle|^2\biggl(
e^{i\omega_{ab}(\tau-\tau')}+e^{-i\omega_{ab}(\tau-\tau')}\biggr)\;,\label{ca}
\end{eqnarray}
\begin{eqnarray}
\chi^A(\tau,\tau')={1\/2}\sum_b|\langle
a|R_2^f(0)|b\rangle|^2\biggl(
e^{i\omega_{ab}(\tau-\tau')}-e^{-i\omega_{ab}(\tau-\tau')}\biggr)\;,\label{xa}
\end{eqnarray}
where $\omega_{ab}=\omega_a-\omega_b$ and the sum extends over a
complete set of atomic states. In order to get the statistical
functions for the field, we consider firstly the two point function
for the field at a nonzero temperature $T=1/\beta$ \cite{QFT}
\begin{eqnarray}
\langle\phi^f(x)\phi^f(x')\rangle_\beta=\textmd{tr}[\rho\phi^f(x)\phi^f(x')]/\textmd{tr}(\rho)\;.
\end{eqnarray}
 From
Eq.~(\ref{phi}), one finds
\begin{eqnarray}
\langle\phi^f(x)\phi^f(x')\rangle_\beta&=&\int_0^\infty
d\omega\int_{-\infty}^\infty dk_y \int_{-\infty}^\infty dk_z
\biggl[\sum_{n=0}(n+1)e^{-n\omega/T}\upsilon_{\omega,k_y,k_z}(x)
\upsilon^\ast_{\omega,k_y,k_z}(x')\nn\\&&+\sum_{n=1}n~e^{-n\omega/T}
\upsilon^\ast_{\omega,k_y,k_z}(x)\upsilon_{\omega,k_y,k_z}(x')\biggr]
\bigg/\sum_{n=0}e^{-n\omega/T}\;.
\end{eqnarray}
Here we assume that there exist an infinite number of Rindler
particles in the thermal bath. We can obtain
\begin{eqnarray}
\langle\phi^f(x)\phi^f(x')\rangle_\beta&=&\int_0^\infty
d\omega\int_{-\infty}^\infty dk_y \int_{-\infty}^\infty
dk_z\biggl[{e^{\omega/T}\/e^{\omega/T}-1}\upsilon_{\omega,k_y,k_z}(x)
\upsilon^\ast_{\omega,k_y,k_z}(x')\nn\\&&+{1\/e^{\omega/T}-1}\upsilon^\ast_{\omega,k_y,k_z}(x)
\upsilon_{\omega,k_y,k_z}(x')\biggr]\;.\label{tp}
\end{eqnarray}
The statistical functions of the field can be calculated using
(\ref{tp}).

\section{Excitation rate in the coaccelerated frame in the unbounded space}

In this section, we will apply the previously developed formalism to
calculate, in the coaccelerated frame, the rate of change of the
mean atomic energy for a uniformly accelerated atom interacting with
a massless free scalar field at finite temperature $T=1/\beta$. In
the Rindler coordinates, the atom is static and its trajectory can
be described  by
\begin{eqnarray}
\tau\;,\ \ \xi=y(\tau)=z(\tau)=0\;.\label{tr1}
\end{eqnarray}
The wave equation for the scalar field in the Rindler coordinates is
given by
\begin{eqnarray}
\biggl[e^{-2a\xi}\biggl({\partial^2\/\partial\tau^2}-{\partial^2\/\partial\xi^2}\biggr)
-{\partial^2\/\partial y^2}-{\partial^2\/\partial
z^2}\biggr]\phi(x)=0\;.
\end{eqnarray}
It can be shown that the orthonormal mode solution with a positive
frequency is
\begin{eqnarray}
\upsilon_{\omega,k_y,k_z}(x)={1\/2\pi^2\sqrt{a}}\sinh^{1/2}\biggl({\pi\omega\/a}\biggr)
K_{i\omega/a}\biggl({1\/a}k_\bot
e^{a\xi}\biggr)e^{ik_yy+ik_zz-i\omega\tau}\;,\label{upsilon}
\end{eqnarray}
where $k_\bot=\sqrt{k_y^2+k_z^2}$ and $K_\nu(x)$ is the Bessel
function of imaginary argument. Now we can evaluate the two point
function (\ref{tp}) for the trajectory (\ref{tr1}), and get
\begin{eqnarray}
\langle\phi^f(x)\phi^f(x')\rangle_\beta&=&{1\/4\pi^4a}\int_0^\infty
d\omega\int_{-\infty}^\infty dk_y \int_{-\infty}^\infty
dk_z\sinh\biggl({\pi\omega\/a}\biggr)K^2_{i\omega/a}\biggl({k_\bot\/a}\biggr)
\nn\\&&\times\biggl({e^{\omega/T}\/e^{\omega/T}-1}e^{-i\omega(\tau-\tau')}
+{1\/e^{\omega/T}-1}e^{i\omega(\tau-\tau')}\biggr)\;.
\end{eqnarray}
With the help of the following integral,
\begin{eqnarray}
\int_{-\infty}^\infty dk_y \int_{-\infty}^\infty
dk_zK^2_{i\omega/a}\biggl({k_\bot\/a}\biggr)={a\pi^2\omega\/\sinh({\pi\omega\/a})}\;,
\end{eqnarray}
we obtain
\begin{eqnarray}
\langle\phi^f(x)\phi^f(x')\rangle_\beta={1\/4\pi^2}\int_0^\infty
d\omega\;\omega\biggl({e^{\omega/T}\/e^{\omega/T}-1}e^{-i\omega(\tau-\tau')}
+{1\/e^{\omega/T}-1}e^{i\omega(\tau-\tau')}\biggr)\;.\label{tp1}
\end{eqnarray}
The statistical functions of the field, (\ref{ct}) and (\ref{xt}),
can now be written
\begin{eqnarray}
C^F_\beta(x,x')={1\/8\pi^2}\int_0^\infty
d\omega\;\omega\;\biggl(1+{2\/e^{\omega/T}-1}\biggr)\biggl(e^{-i\omega
(\tau-\tau')}+e^{i\omega(\tau-\tau')}\biggr)\;,\label{cf1}
\end{eqnarray}
\begin{eqnarray}
\chi^F_\beta(x,x')={1\/8\pi^2}\int_0^\infty
d\omega\;\omega\;\biggl(e^{-i\omega
(\tau-\tau')}-e^{i\omega(\tau-\tau')}\biggr)\;.\label{xf1}
\end{eqnarray}
Note that $\chi^F_\beta(x,x')$ has no temperature dependence and
agrees with the linear susceptibility in vacuum. This is a result of
the fact that only the field commutator appears in (\ref{xt}) and
the linear susceptibility thus does not depend on the state of the
field. With a substitution $u=\tau-\tau'$, the contributions
Eq.~(\ref{tf}) and Eq.~(\ref{rr}) to the rate of change of the
atomic energy can be evaluated to get
\begin{eqnarray}
\biggl\langle{dH_A(\tau)\/d\tau}\biggr\rangle_{TF}^R&=&-{\mu^2\/8\pi^2}\sum_b\omega_{ab}|\langle
a|R_2^f(0)|b\rangle|^2\int_0^\infty
d\omega\;\omega\;\biggl(1+{2\/e^{\omega/T}-1}\biggr)\nn\\&&\times\int_0^\infty
du(e^{-i\omega u}+e^{i\omega
u})(e^{-i\omega_{ab}u}+e^{i\omega_{ab}u})
\end{eqnarray}
\begin{eqnarray}
\biggl\langle{dH_A(\tau)\/d\tau}\biggr\rangle_{RR}^R&=&-{\mu^2\/8\pi^2}\sum_b\omega_{ab}|\langle
a|R_2^f(0)|b\rangle|^2\int_0^\infty
d\omega\;\omega\nn\\&&\times\int_0^\infty du(e^{-i\omega
u}-e^{i\omega u})(e^{-i\omega_{ab}u}-e^{i\omega_{ab}u})\;.
\end{eqnarray}
Here we have extended the range of integration to infinity for
sufficiently long times $\tau-\tau_0$. After the evaluation of the
integrals we obtain
\begin{eqnarray}
\biggl\langle{dH_A(\tau)\/d\tau}\biggr\rangle_{TF}^R&=&-{\mu^2\/2\pi}
\biggl[\sum_{\omega_a>\omega_b} \omega_{ab}^2|\langle
a|R_2^f(0)|b\rangle|^2\biggl({1\/2}+{1\/e^{\omega_{ab}/T}-1}\biggr)
\nn\\&&-\sum_{\omega_a<\omega_b}\omega_{ab}^2|\langle
a|R_2^f(0)|b\rangle|^2\biggl({1\/2}+{1\/e^{|\omega_{ab}|/T}-1}\biggr)\biggr]\label{tf1}
\end{eqnarray}
for the contribution of the thermal fluctuations to the rate of
change of atomic excitation energy, and
\begin{eqnarray}
\biggl\langle{dH_A(\tau)\/d\tau}\biggr\rangle_{RR}^R=-{\mu^2\/2\pi}\biggl(\sum_{\omega_a>\omega_b}
{1\/2}\omega_{ab}^2|\langle
a|R_2^f(0)|b\rangle|^2+\sum_{\omega_a<\omega_b}{1\/2}\omega_{ab}^2|\langle
a|R_2^f(0)|b\rangle|^2\biggr)\label{rr1}
\end{eqnarray}
for that of radiation reaction. A comparison with the spontaneous
excitation rate of an accelerated atom calculated in the inertial
frame \cite{Audretsch94} shows that the contribution of radiation
reaction to the rate of change of the atomic energy  is the same as
that calculated in the inertial frame, and thus is observer
independent and temperature independent. While the contributions of
vacuum fluctuations are equal only when the temperature equals the
FDU value, i.e., $T=a/2\pi$. Hence the same total rate of change of
the atomic energy can be obtained both in the inertial and
coaccelerated frames, only when we assume a thermal state of Rindler
particles at a temperature $T=a/2\pi$ to a coaccelerated observer
for what appears to be a vacuum state to an inertial observer. This
is the FDU effect.

\section{a space-time with a reflecting plane boundary}

It is well known that the presence of boundaries modifies the
quantum fluctuations of fields both in a vacuum state and a thermal
state and can lead to a lot of novel effects, such as the Casimir
effect \cite{Casimir}, the light-cone fluctuations \cite{YF}, the
Brownian (random) motion of test particles caused by quantum
fluctuations in vacuum and quantum fluctuations at finite
temperature \cite{Yu04}, and so on. Therefore, the investigation of
spontaneous excitation rate of accelerated atoms interacting with
the scalar field in the presence of a reflecting plane boundary is
an interesting issue and has been carried out in the inertial frame
\cite{H. Yu05}. The results obtained there show that,  with the
presence of a reflecting boundary, a uniformly accelerated atom in a
vacuum does {\it not} have to behave as if it were static in a
thermal bath at a temperature $T=a/2\pi$,  in the sense that the
spontaneous excitation rate is changed,  by the presence of the
boundary, in such a way that it is different from what one would
expect for an inertial atom in a thermal bath in the same space with
the boundary. We will, however, show that the same excitation rate
can be obtained in the coaccelerated frame assuming the FDU effect.

We assume that a perfectly reflecting boundary for the scalar field
is located at $z=0$ in space, the atom is being uniformly
accelerated in the $x$ direction with a proper acceleration $a$ at a
distance $z$ from the boundary, and the whole system is in a thermal
bath at a temperature $T=1/\beta$. The atom's trajectory is now
described in the Rindler coordinates by
\begin{eqnarray}
\tau\;,\ \ \xi=y(\tau)=0\;,\ \ z(\tau)=z\;.\label{trajectory}
\end{eqnarray}
To satisfy the Dirichlet boundary condition, $\phi(x)|_{z=0}=0$,
the normalized Rindler mode function with a positive frequency for
the scalar field becomes
\begin{eqnarray}
\upsilon_{\omega,k_y,k_z}(x)={1\/\pi^2\sqrt{2a}}\sinh^{1/2}\biggl({\pi\omega\/a}\biggr)
K_{i\omega/a}\biggl({1\/a}k_\bot
e^{a\xi}\biggr)\sin(k_zz)e^{ik_yy-i\omega\tau}\;.
\end{eqnarray}
From Eq.~(\ref{tp}) we can calculate the two-point function of the
field at a finite temperature $T=1/\beta$ for the trajectory
(\ref{trajectory}) to get
\begin{eqnarray}
\langle\phi^f(x)\phi^f(x')\rangle_\beta&=&{1\/4\pi^4a}\int_0^\infty
d\omega\int_{-\infty}^\infty dk_y \int_{-\infty}^\infty
dk_z\sinh\biggl({\pi\omega\/a}\biggr)K^2_{i\omega/a}\biggl({k_\bot\/a}\biggr)[1-\cos(2k_zz)]
\nn\\&&\times\biggl({e^{\omega/T}\/e^{\omega/T}-1}e^{-i\omega(\tau-\tau')}
+{1\/e^{\omega/T}-1}e^{i\omega(\tau-\tau')}\biggr)\;.
\end{eqnarray}
With the help of the following integral
\begin{eqnarray}
\int_{-\infty}^\infty dk_y \int_{-\infty}^\infty
dk_zK^2_{i\omega/a}\biggl({k_\bot\/a}\biggr)[1-\cos(2k_zz)]={a\pi^2\/\sinh({\pi\omega\/a})}
\biggl[\omega-{\sin({2\omega\sinh^{-1}(az)\/a})\/2z\sqrt{1+a^2z^2}}\biggr]\;,
\end{eqnarray}
we find
\begin{eqnarray}
\langle\phi^f(x)\phi^f(x')\rangle_\beta&=&{1\/4\pi^2}\int_0^\infty
d\omega\biggl[\omega-{\sin({2\omega\sinh^{-1}(az)\/a})\/2z\sqrt{1+a^2z^2}}\biggr]
\nn\\&&\times\biggl({e^{\omega/T}\/e^{\omega/T}-1}e^{-i\omega(\tau-\tau')}
+{1\/e^{\omega/T}-1}e^{i\omega(\tau-\tau')}\biggr)\;.\nn\\\label{tp2}
\end{eqnarray}
From the general form (\ref{ct}) and (\ref{xt}), we can obtain the
corresponding statistical functions of the field
\begin{eqnarray}
C^F_\beta(x,x')={1\/8\pi^2}\int_0^\infty
d\omega\biggl[\omega-{\sin({2\omega\sinh^{-1}(az)\/a})\/2z\sqrt{1+a^2z^2}}\biggr]
\biggl(1+{2\/e^{\omega/T}-1}\biggr)\biggl(e^{-i\omega
(\tau-\tau')}+e^{i\omega(\tau-\tau')}\biggr)\;,\nonumber\\\label{cf2}
\end{eqnarray}
\begin{eqnarray}
\chi^F_\beta(x,x')={1\/8\pi^2}\int_0^\infty
d\omega\biggl[\omega-{\sin({2\omega\sinh^{-1}(az)\/a})\/2z\sqrt{1+a^2z^2}}\biggr]
\biggl(e^{-i\omega(\tau-\tau')}-e^{i\omega(\tau-\tau')}\biggr)\;.\label{xf2}
\end{eqnarray}
Again $\chi^F_\beta(x,x')$ has no temperature dependence as
expected, it however does rely on the acceleration.  With a
substitution $u=\tau-\tau'$ and extending the range of integration
to infinity for sufficiently long times $\tau-\tau_0$, we  find the
contributions of thermal fluctuations (\ref{tf}) and radiation
reaction (\ref{rr}) to the rate of change of the atomic energy,
\begin{eqnarray}
\biggl\langle{dH_A(\tau)\/d\tau}\biggr\rangle_{TF}^R&=&-{\mu^2\/2\pi}\biggl[
\sum_{\omega_a>\omega_b}{1\/2}\omega_{ab}^2|\langle
a|R_2^f(0)|b\rangle|^2f(\omega_{ab},a,z)\biggl({1\/2}+{1\/e^{\omega_{ab}/T}-1}\biggr)
\nn\\&&-\sum_{\omega_a<\omega_b}{1\/2}\omega_{ab}^2|\langle
a|R_2^f(0)|b\rangle|^2f(\omega_{ab},a,z)\biggl({1\/2}+{1\/e^{|\omega_{ab}|/T}-1}\biggr)\biggr]\label{tf2}\;,
\end{eqnarray}
and
\begin{eqnarray}
\biggl\langle{dH_A(\tau)\/d\tau}\biggr\rangle_{RR}^R&=&-{\mu^2\/2\pi}\biggl[
\sum_{\omega_a>\omega_b}{1\/2}\omega_{ab}^2|\langle
a|R_2^f(0)|b\rangle|^2f(\omega_{ab},a,z)\nn\\&&+\sum_{\omega_a<\omega_b}{1\/2}\omega_{ab}^2|\langle
a|R_2^f(0)|b\rangle|^2f(\omega_{ab},a,z)\biggr]\label{rr2}\;,
\end{eqnarray}
where
\begin{eqnarray}
f(\omega_{ab},a,z)=1-{1\/2\omega_{ab}z\sqrt{1+a^2z^2}}\sin\biggl({2\omega_{ab}\sinh^{-1}(az)\/a}\biggr)\;.
\end{eqnarray}

Comparing the above results with Eqs.~(\ref{tf1}) and (\ref{rr1}),
one can see that the function $f(\omega_{ab},a,z)$ gives the
modification induced by the presence of the boundary.  Letting
$T\rightarrow0$, we obtain the contribution of vacuum fluctuations
\begin{eqnarray}
\biggl\langle{dH_A(\tau)\/d\tau}\biggr\rangle_{TF}^R\bigg|_{T\rightarrow0}
&=&-{\mu^2\/2\pi}\biggl[\sum_{\omega_a>\omega_b}{1\/2}\omega_{ab}^2|\langle
a|R_2^f(0)|b\rangle|^2f(\omega_{ab},a,z)\nn\\&&-\sum_{\omega_a<\omega_b}{1\/2}\omega_{ab}^2|\langle
a|R_2^f(0)|b\rangle|^2f(\omega_{ab},a,z)\biggr]\label{tf2,t=0}\;,
\end{eqnarray}
Adding up the contributions of vacuum fluctuations (\ref{tf2,t=0})
and radiation reaction (\ref{rr2}) to the Rindler atom yields the
total spontaneous excitation rate,
\begin{eqnarray}
\biggl\langle{dH_A(\tau)\/d\tau}\biggr\rangle_{tot}^R\bigg|_{T\rightarrow0}
=-{\mu^2\/2\pi}\sum_{\omega_a>\omega_b}\omega_{ab}^2|\langle
a|R_2^f(0)|b\rangle|^2f(\omega_{ab},a,z)\;.
\end{eqnarray}
This indicates that if no FDU thermal bath were assumed to exist,
the Rindler atom would be stable in its ground state with no
excitation. Yet this is different from the behavior of an inertial
atom (cf. Eq.~(23) in Ref.~\cite{H. Yu05}), because of the
appearance of the acceleration $a$ in function $f(\omega_{ab},a,z)$,
revealing that without assuming the FDU effect what an coaccelerated
observer would see will be different from an inertial observer. This
is in contrast to the case without the presence of the boundary.

Now, let us recall the spontaneous excitation rate of a uniformly
accelerated atom in the inertial frame, with the presence of a plane
boundary \cite{H. Yu05}. The contributions of vacuum fluctuations
and radiation reaction are
\begin{eqnarray}
\biggl\langle{dH_A(\tau)\/d\tau}\biggr\rangle_{VF}^{Inertial}&=&-{\mu^2\/2\pi}\biggl[
\sum_{\omega_a>\omega_b} \omega_{ab}^2|\langle
a|R_2^f(0)|b\rangle|^2f(\omega_{ab},a,z)\biggl({1\/2}+{1\/e^{2\pi\omega_{ab}/a}-1}\biggr)
\nn\\&&-\sum_{\omega_a<\omega_b}\omega_{ab}^2|\langle
a|R_2^f(0)|b\rangle|^2f(\omega_{ab},a,z)
\biggl({1\/2}+{1\/e^{2\pi|\omega_{ab}|/a}-1}\biggr)\biggr]\;,\label{vfi2}
\end{eqnarray}
and
\begin{eqnarray}
\biggl\langle{dH_A(\tau)\/d\tau}\biggr\rangle_{RR}^{Inertial}
&=&-{\mu^2\/2\pi}\biggl[\sum_{\omega_a>\omega_b}{1\/2}\omega_{ab}^2|\langle
a|R_2^f(0)|b\rangle|^2f(\omega_{ab},a,z)\nn\\&&+\sum_{\omega_a<\omega_b}{1\/2}\omega_{ab}^2|\langle
a|R_2^f(0)|b\rangle|^2f(\omega_{ab},a,z)\biggr]\label{rri2}\;.
\end{eqnarray}
 A comparison with Eq.~(\ref{tf2}) and Eq.~(\ref{rr2}) shows that,
the same as in the unbounded case, the contribution of radiation
reaction is again observer independent, however, only when we assume
a thermal bath of Rindler particles at the temperature $T=a/2\pi$
for the coaccelerated observer, is Eq.~(\ref{tf2}) equal to
Eq.~(\ref{vfi2}) and the two observers agree on the total rate of
the change of the atomic energy for the accelerated atom. Let us
note that, due to the appearance of $f(\omega_{ab},a,z)$, the total
spontaneous excitation rate of a uniformly accelerated atom in the
presence of a reflecting boundary deviates from the pure thermal
rate in the unbounded case. However, this discrepancy does not imply
that the exact final thermal equilibrium is not
achieved~\cite{GP04}. As a matter of fact, since function
$f(\omega_{bd},z,a)$ is an even function of $\omega_{bd}$, one can
show, by the same argument as that in Ref~\cite{GP04}, that exact
thermal equilibrium will be established at the FDU temperature
$T_{FDU}=a/2\pi$, even though the accelerated atom radiates and
absorbs differently from an inertial atom immersed in the thermal
bath.

\section{conclusions}

In conclusion, from the point of view of the coaccelerated frame,
we have examined the spontaneous excitation of  a uniformly
accelerated two-level atom interacting with a massless quantum
scalar field both in a free space and in a space with a reflecting
plane boundary and separately calculated the contributions of
thermal fluctuations and radiation reaction to the rate of change
of  the mean atomic energy, assuming a thermal bath of Rindler
scalar particles at a finite temperature $T$. Our results show
that the same spontaneous excitation rate for a uniformly
accelerated atom in the Minkowski vacuum obtained in the inertial
frame can be recovered in the coaccelerated frame ONLY when
assuming a thermal bath at the Fulling-Davies-Unruh temperature
$T_{FDU}=a/2\pi$  for what appears to be the Minkowski vacuum to
the inertial observer.

As an example of a different physical process, our results give
further support to endeavors already made by other authors in other
different contexts ~\cite{Matsas01,Higuchi92,Hai Ren,Danilo} in
clarifying the confusion on what the Fulling-Davies-Unruh effect
means.  The Fulling-Davies-Unruh effect implies that a uniformly
accelerated atom (observer) interprets as a thermal bath of Rindler
particles at the temperature $T_{FDU}=a/2\pi$ what an inertial
observer sees as a vacuum devoid of particles, or  in different
words, the physical observables calculated in the inertial frame can
be recovered in the coaccelerated frame by using Fulling's
quantization in conjunction with the fact that the Minkowski vacuum
is a thermal state of Rindler particles at a temperature
$T_{FDU}=a/2\pi$.

\begin{acknowledgments}
This work was supported in part  by the National Natural Science
Foundation of China  under Grants No.10375023 and No.10575035, and
the Program for New Century Excellent Talents in University (NCET,
No. 04-0784).
\end{acknowledgments}

\end{document}